# 3D Monte Carlo Simulation of Light Distribution in Mouse Brain in Quantitative Photoacoustic Computed Tomography

Yuqi Tang, Junjie Yao*

Department of Biomedical Engineering, Duke University, Durham, NC, USA

junjie.yao@duke.edu

**Abstract:** Photoacoustic computed tomography (PACT) detects light-induced ultrasound waves to reconstruct the optical absorption contrast of the biological tissues. Due to its relatively deep penetration (several centimeters in soft tissue), high spatial resolution, and inherent functional sensitivity, PACT has great potential for imaging mouse brains with endogenous and exogenous contrasts, which is of immense interest to the neuroscience community. However, conventional PACT either assumes homogenous optical fluence within the brain or uses a simplified attenuation model for optical fluence estimation. Both approaches underestimate the complexity of the fluence heterogeneity and can result in poor quantitative imaging accuracy. To optimize the quantitative performance of PACT, we explore for the first time 3D Monte Carlo simulation to study the optical fluence distribution in a complete mouse brain model. We apply the MCX Monte Carlo simulation package on a digital mouse (Digimouse) brain atlas that has complete anatomy information. To evaluate the impact of the brain vasculature on light delivery, we also incorporate the whole-brain vasculature in the Digimouse atlas. The simulation results clearly show that the optical fluence in the mouse brain is heterogeneous at the global level and can decrease by a factor of five with increasing depth. Moreover, the strong absorption and scattering of the brain vasculature also induce the fluence disturbance at the local level. Our results suggest that both global and local fluence heterogeneity contributes to the reduced quantitative accuracy of the reconstructed PACT images of mouse brain.

**Key words:** Quantitative photoacoustic imaging; photoacoustic computed tomography; 3D Monte Carlo simulation; mouse brain imaging; digital mouse brain; whole-brain vasculature; optical fluence distribution

# 1 Introduction

Small animal neuroimaging has gained increasing interest in the past decades. This appeal is motivated by the progress in the transgenic manipulation of small animals, especially mice, as models of human brain diseases and pathological conditions. As mouse models share many genes, physiological processes, and disease loci with humans, investigation of these models helps improve the understanding, prevention, diagnosis, and treatment of human diseases. A variety of imaging modalities have been applied for small animal brain imaging, including magnetic resonance microscopy (MRM), positron emission tomography (PET), single-photon emission computed tomography (SPECT), ultrasound (US) imaging, and multiphoton microscopy. Using these imaging modalities, great advancements have been made in understanding brain function and neurological pathologies, yet major technical challenges remain unsolved.

MRM, or high-resolution magnetic imaging (MRI), can visualize neuroanatomical structures at resolutions <100 μm in at least one dimension. Yet tradeoffs exist between imaging time and spatial resolution, and long scanning times are required for better signal-to-noise ratio (SNR) [1]. PET relies on radioactive tracer molecules and provides quantitative measure in living organisms, including regional cerebral glucose utilization, oxygen metabolism, and cerebral blood flow. However, the application of small animal PET is limited by its low availability, high ionizing exposure, relatively high cost, and low spatial resolution [2]. Like PET, SPECT is a nuclear imaging technique, and can reach a resolution of 0.35 mm, perform longitudinal studies, and distinguish among multiple radioisotopes [3]. However, SPECT also suffers from the potential high ionizing exposure. US imaging is nonionizing and derives contrast from the mechanical properties of the tissue. US imaging can provide tissue morphology and map blood flow with resolution as high as 48 μm in rodent brain with the aid of microbubbles. However, US imaging cannot provide information about the brain's neuroactivities so far. Multiphoton microscopy (two-photon (2PM) or three-photon microscopy), which provides micron-scale resolution and readily reveals brain cortical organization and function, is the most widely used optical technology for *in vivo* mouse brain imaging, at depths up to 1.5 mm using contrast agents that are bright/biocompatible and absorbing at longer wavelengths [4]. Optical coherence tomography (OCT) has also been explored for mouse brain imaging, although it lacks the cellular and molecular specificity [5]. However, both multiphoton microscopy and OCT lack the penetration depth to study the mouse brain beyond the cortical layer.

In the past several decades, photoacoustic (PA) imaging has gradually attracted attention for rodent brain imaging. PA imaging is a nonionizing hybrid imaging modality that physically combines light and sound. In typical PA imaging, short laser pulse illuminates the target, and the incident photons are subsequently absorbed by the chromophores in the target. The absorption leads to a transient increase in temperature and induces a local pressure rise, resulting in the emission of broadband ultrasonic waves [6]. Hemoglobin is one of the most commonly used chromophores in PA imaging as its two forms, oxygenated and deoxygenated, have strong optical absorption in the visible and near-infrared spectral range. With multispectral PA imaging, functional information such as blood oxygenation can also be estimated. Depending upon the configuration of optical illumination and ultrasound detection, PA imaging can be further classified into two major types:

PA microscopy (PAM) and PA computed tomography (PACT). While PAM relies on tightly or weakly focused light for excitation and uses a single element focused ultrasonic transducer for direct image formation. It provides high spatial resolution in ballistic and quasidiffusive regimes to image rodent cortex [7,8][9–11]. PACT relies on homogenized widefield illumination and the photons can reach the diffusive regime. The PA signals are acquired with an ultrasonic transducer array, and the images are formed with an inverse algorithm [8,12]. PACT has been widely used in rodent whole brain imaging as it allows monitoring of the brain hemodynamics with balanced spatiotemporal resolution and field of view. Based on the estimated hemoglobin concentration and change in blood oxygenation, neuronal activity can be studied through the neurovascular coupling. PACT has been used to study epilepsy or Alzheimer's disease [13–15], glioblastoma tumor growth [16], and circulating melanoma cancer cells [17]. Moreover, PACT is highly compatible with commercially available ultrasound (US) imaging systems [18–24].

Most of the functional information provided by PACT, such as blood oxygenation, heavily relies on the estimation of the chromophore concentration. Based on the imaging formation mechanism, the amplitude of the initial PA wave pressure mainly depends on three parameters: the absorption coefficient of the chromophore, the local optical fluence, and the Gruneisen parameter [6,25]. In most *in vivo* cases, the PA signal amplitude is assumed to be directly proportional to the chromophore concentration and does not consider the optical fluence heterogeneity across the region of interest. However, as the brain is a strongly scattering organ and functional regions possess different optical properties, the PA signal strength could reflect different combinations of local optical fluence and chromophore concentrations. Different methods have been reported in order to overcome this limitation and improve the quantitative fidelity of the reconstructed PA images, including adaptive filtered back-projection and model-based reconstruction with fluence distribution estimation [26,27]. However, these methods have difficulty for *in vivo* use, especially in mouse brain tissues that have dense structures.

Therefore, to better understand the impact of optical fluence heterogeneity in mouse brain on the PA imaging and potentially correct it, we propose to apply Monte Carlo (MC) method to simulate the light propagation and resultant optical fluence distribution inside the mouse brain. The MC method constructs a stochastic model to determine the expected light propagation and is the gold standard for studying fluence distribution in biological tissue. It is relatively easy to implement and widely accepted as an accurate method by which to simulate light propagation in tissues [28–30]. The MC method has been utilized in PA imaging for simulating imaging depth at different wavelengths, optimizing experimental setup for maximum and homogenous light delivery, and studying the optical fluence distribution inside human infant brain [31–34]. So far, the light propagation and fluence distribution inside mouse brain remains to be studied for quantitative PACT. In this paper, we have investigated the optical fluence distribution inside the mouse brain while taking blood vessels into account, and its subsequent impact on the reconstructed PA images.

## 2 Methods

### 2.1 *Digital mouse brain model*

As blood vessels are highly absorbing and scattering compared to other brain tissues, even though they make up only ~5% of the total brain volume, their impact on optical fluence distribution cannot be omitted in *in vivo* experiments. A hybrid model was thus constructed with blood vessels identified separately from other brain tissues. Due to the lack of a model that identifies both anatomical structure and blood vessel distribution, a hybrid mouse brain model with vasculature identified was created using two published data sets: the Digimouse atlas developed by Dogdas *et al.* and light sheet fluorescent microscopy (LSFM) images of mouse-brain vasculature acquired by Di Giovanna *et al.* (Figure 1(a)) [35,36]. Both data sets are volumetric. The Digimouse atlas identifies mouse brain functional regions, such as olfactory bulbs and external cerebrum, as well as anatomical structures, such as skin and skull, all of which are included in the tetrahedral mesh for simulation (Figure 1(b)). LSFM mouse-brain vasculature images provide the blood vessel population density in different parts of the mouse brain. The hybrid brain model with vasculature identified (Digimouse-vasculature atlas, in short) was created by manually registering the LSFM mouse-brain vasculature images with Digimouse atlas through resizing, shifting, and thresholding (Figure 1(c-e)).

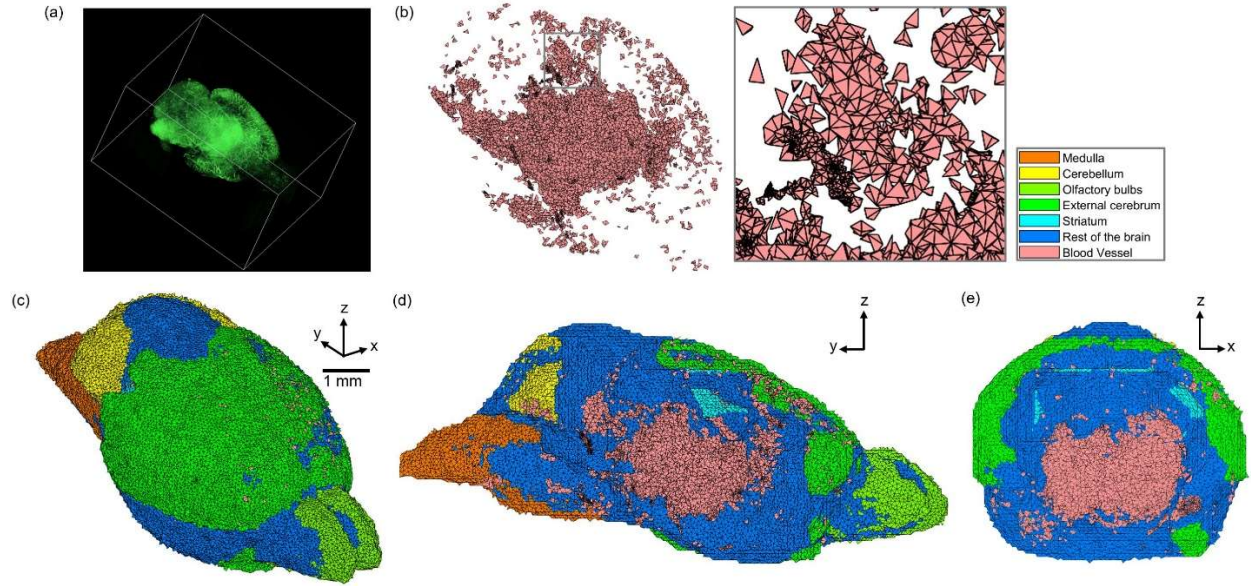

Figure 1. Digital mouse brain model with blood vasculature incorporated for 3D Monte Carlo simulation. The blood vasculature information was obtained from light sheet fluorescence microscopy (LSFM) imaging. The LSFM image was scaled and shifted manually to register with Digimouse model. The merged model was then converted to tetrahedral mesh. (a) 3D rendering of LSFM blood vasculature image. (b) Blood vasculature distribution in the model, assuming whole blood with a 45% hematocrit. The thalamus and cortex have a relatively high vessel distribution density. (c) Overview of the merged model. Different colors represent different regions identified in brain. Skin and skull are not shown. (d) A representative sagittal plane around the middle of the brain. (e) A representative transversal plane around the bregma.

### 2.2 *3D Monte Carlo Modeling Software*

MMCLAB, the native MEX version of mesh-based Monte Carlo (MMC) photon simulation software for MATLAB, was used for our simulation [37,38]. Different from existing MC software designed for layered or voxel-based media, MMC can represent a complex domain using a volumetric mesh. It can utilize a tetrahedral mesh to model a complex anatomical structure and has been shown to be more accurate and computationally efficient than the conventional MC methods [39,40]. In MMC, anatomical structures in the model are identified and assigned with optical properties, including the scattering coefficient ($\mu_s$), absorption coefficient ($\mu_a$), anisotropy (g), and refractive index (n). The model is then subsequently converted to tetrahedral mesh with iso2mesh MATLAB toolbox [41]. Other configurations, including light source type, location, illumination direction, and number of photons, are predefined before the simulation. The MMC software then simulates photon propagation from the predefined light source through the volume consisted of tetrahedral voxels, and outputs optical fluence distribution normalized to the initial optical energy launched into the simulation volume.

## 2.3 3D Monte Carlo Model Construction

In this work, to study the potential effect of blood vasculature on optical fluence distribution in mouse brain, we performed 3D MC simulations on mouse brain models with and without integrated blood vasculature. Both the Digimouse and Digimouse-vasculature atlas were converted into tetrahedral mesh using the iso2mesh toolbox [41]. To validate the hybrid Digimouse-vasculature mesh, we estimated the cerebral blood volume (CBV) percentage by calculating the element number ratio between blood vessels and other brain tissues [42]. The estimated CBV ratio was 5.9%, with relatively high blood vessel population density in cortex, hypothalamus, and thalamus. Though the two data sets are from two different sources and cannot be perfectly registered, the overall blood vessel population density within the hybrid model is consistent with the reported literature values [42].

## 2.4 3D Monte Carlo Modeling Parameters

Table 1 Optical properties of brain tissues for 3D Monte Carlo simulation, approximated from human data.

| Tissue type | Absorption coefficient, $\mu_a$ (mm$^{-1}$) | Scattering coefficient, $\mu_s$ (mm$^{-1}$) | Anisotropy, g | Refractive index, n | Ref. |
|---|---|---|---|---|---|
| Scalp | 0.0191 | 6.6 | 0.9 | 1.4 | [43,44] |
| Skull | 0.0136 | 8.6 | 0.9 | 1.4 | [43,44] |
| Brain tissue* | 0.0186 | 11.1 | 0.9 | 1.4 | [43,44] |
| Blood vessel | 0.36 | 59.6 | 0.97 | 1.4 | [45,46] |

*Includes medulla, cerebellum, olfactory bulbs, external cerebrum, stratum, and rest of the brain.*

In PACT of mouse brains, a linear array transducer is commonly used due to its wide availability and high compatibility with commercial ultrasound scanners. Figure 2(a) shows a schematic of a common PACT brain imaging setup used in our simulations. The linear array transducer such as L22-14v (Verasonics Inc., Kirkland, WA), is placed on top of the mouse head with coupling gel to image the sagittal plane (Figure 2(b)). A linear illumination pattern is chosen because it can provide relatively homogenous optical fluence at the imaging plane. Such illumination is achieved either by mounting optical fiber bundles on both sides of the transducer array or by guiding a free-space linear light beam directly onto the sample surface. In our simulation, the linear light sources were placed outside the model, and each source provided a uniform fluence at the scalp surface with an area of 2 x 6 mm$^2$. The distance between the light incident position and the center of the transducer imaging plane was 3 mm (Figure 2(a)). As suggested by Sower *et al.*, the illumination angle was set to 45-deg to achieve highest fluence at the imaging plane [32]. The transducer and light sources were immersed in water as the background medium. We assume the background medium was non-scattering and non-absorbing and had no refractive index mismatch with the mouse model.

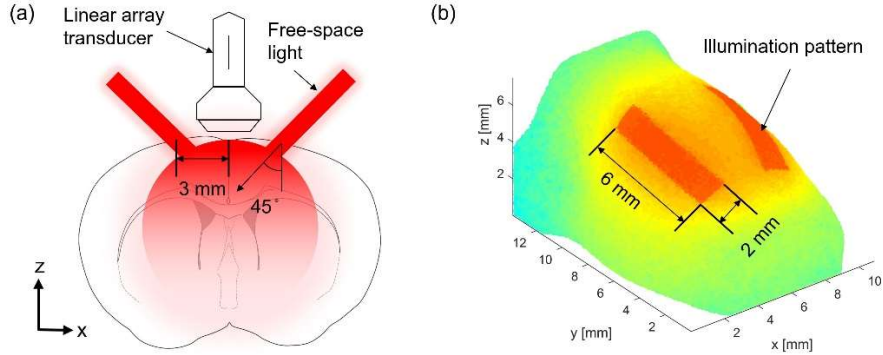

Figure 2: Schematics of simulation system setup. (a) Illumination is provided from two sides of the transducer with a 45-deg angle, while the transducer is placed on top of the mouse head. (b) Illumination pattern on the scalp surface.

Near-infrared (NIR) light is preferred for deep brain imaging due to the relatively low optical scattering from tissues and low optical absorption from water. Using hemoglobin as the endogenous contrast, NIR light can provide relatively deep penetration [6,7]. Here, we chose 1064 nm as the excitation wavelength, which can be readily generated by commonly available Nd:YAG lasers. In our simulations, two different meshes were simulated at 1064 nm using the optical parameters listed in Table 1.

Although further studies are required, Hoshi *et al.* have suggested that the optical properties of rodent and human tissues are comparable [47]. Because the optical properties of mouse brain tissue at 1064 nm are not available, we used the optical properties of human tissues based on the work by Jacques *et al* [43]. The main tissue types in the hybrid mouse model include scalp, skull, eye, brain (with several functional regions), muscles, glands, and blood vessels [43]. The optical properties of glands were approximated to those of muscles. Since the lipid makes up 40% of the brain's dry weight, the absorption coefficient of the brain tissue was approximated to that of lipid [48]. As it is computationally challenging to assign each individual blood vessel element with different optical properties, we assumed an average hematocrit of 45% and an average blood oxygenation (sO$_2$) of 85% [45,49,50]. Here, we need to note that the reported optical properties of the brain tissues in the literatures might include the contributions from the blood vessels, depending on the preparation of the specimen for the optical measurement. Nevertheless, we consider the contribution of blood vessels, especially on the µ$_s$, to be negligible as the average CBV ratio in the brain is only 5% [42].

**2.4** *k-Wave Modeling Parameters*

Quantitative PA imaging relies on accurate estimation of optical fluence inside the tissue, but most of the PA imaging reconstruction algorithms assume constant fluence distribution; therefore, we investigated the effect of inhomogeneous illumination on reconstructed images using the k-Wave toolbox [51]. With the same experimental setup as shown in Figure 2(a), a 256 × 256 2D k-Wave grid was created to cover a region of 12.8 × 12.8 $mm^2$. A group of blood vessels were used to simulate the deep brain vasculature, and the initial PA pressure was simulated based on the fluence distribution obtained from the 3D MC simulation results. The simulated PA signals were then band-pass filtered to assimilate the experimental ultrasound transducer with limited detection bandwidth, and then reconstructed with a time-reversal based method or a delay-and-sum based method [51,52]. The other brain tissues were assumed to be acoustically homogenous, with constant tissue density and zero acoustic attenuation.

## 3 Results

### 3.1 *Overall 3D optical fluence distribution in mouse brain*

The simulated optical fluence maps of several transverse and sagittal planes at selected locations are shown in Figure 3 and Figure 4, respectively, from models with and without blood vessels. As expected, due to the strong optical scattering of brain tissues, the light is largely diffused after propagating just a few millimeters. In the imaging plane, the optical fluence remains relatively homogenous within the cortex region (1 mm deep) and decreases by a factor of 5 at 5 mm depth. Outside the imaging plane, the optical fluence is much stronger near the incident location and decreases by a factor of 10 within 2–3 mm propagation. The high optical fluence outside the imaging plane can generate strong out-of-plane signals and result in image artifacts, yet currently there are very few alternative illumination schemes available for linear array transducers.

### 3.2 *The influence of blood vessels on optical fluence distribution*

Moreover, the simulation results show that blood vessels have clear influence on the optical fluence distribution. Macroscopically, the optical fluence maps are similar in MC models with and without incorporating blood vessels as shown in Figures3 (d-i) and Figures 4(d-i), but the existence of blood vessels clearly changes the local optical fluence at most of the brain regions (Figures 3(j-l) and Figures 4(j-l)). We analyzed the optical fluence as a function of depth at six locations, y = 4, 5, 6, 7, 8, 9 mm, along the sagittal suture (x = 6 mm) (Figure 5). At each depth of the selected location, the total optical fluence within a 0.2 × 0.2 $mm^2$ area was plotted. The optical fluence remained relatively flat for the first 1-2 mm beneath the scalp surface, which is mainly the cortex layer, regardless of the existence of blood vessels. However, the optical fluence in the cortex is significantly lower with the blood vessels. Such difference gradually diminishes in the deeper brain beyond 4 mm. As most of the blood vessel elements are concentrated in the cortex, hypothalamus, and thalamus, the strong optical attenuation of blood vessels reduces the local optical fluence. Since the cortex is closer to the illumination location and thus has a higher optical fluence, the fluence reduction in fluence is more pronounced than other blood-rich regions. In addition, even in deep brain, where the blood vessel population becomes denser in the model, less light is available and thus the effect of the blood vessels on the fluence is less obvious.

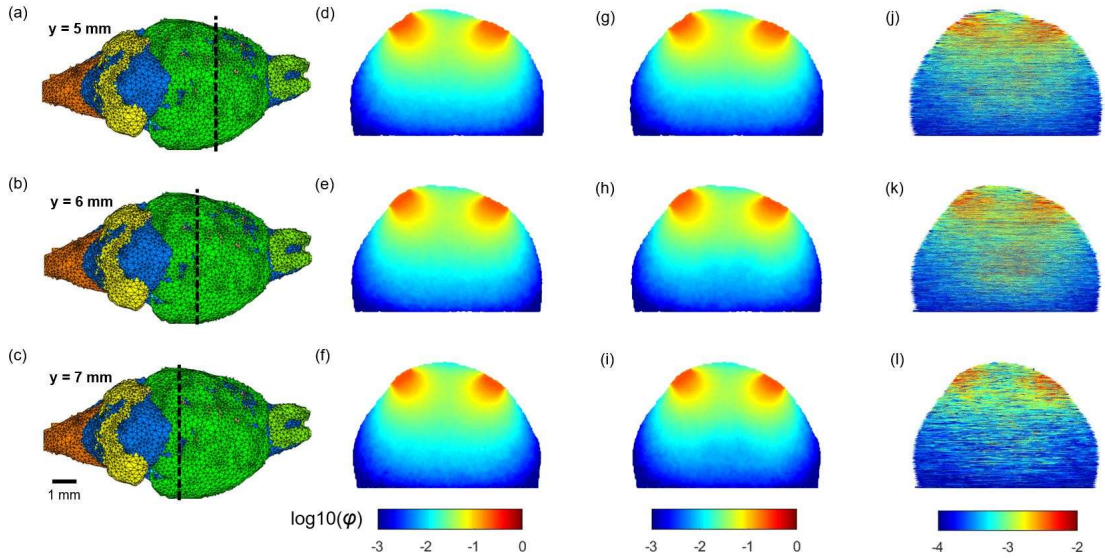

Figure 3: Optical fluence distribution maps of representative transverse planes in 3D MC models with and without blood vessels. (a-c) Black dashed lines indicate the locations of the transverse planes. (d-f) Normalized optical fluence maps without blood vessels shown in log scale. (g-i) Optical fluence maps with blood vessels. (j-l) Optical fluence difference between (d-f) and (g-i).

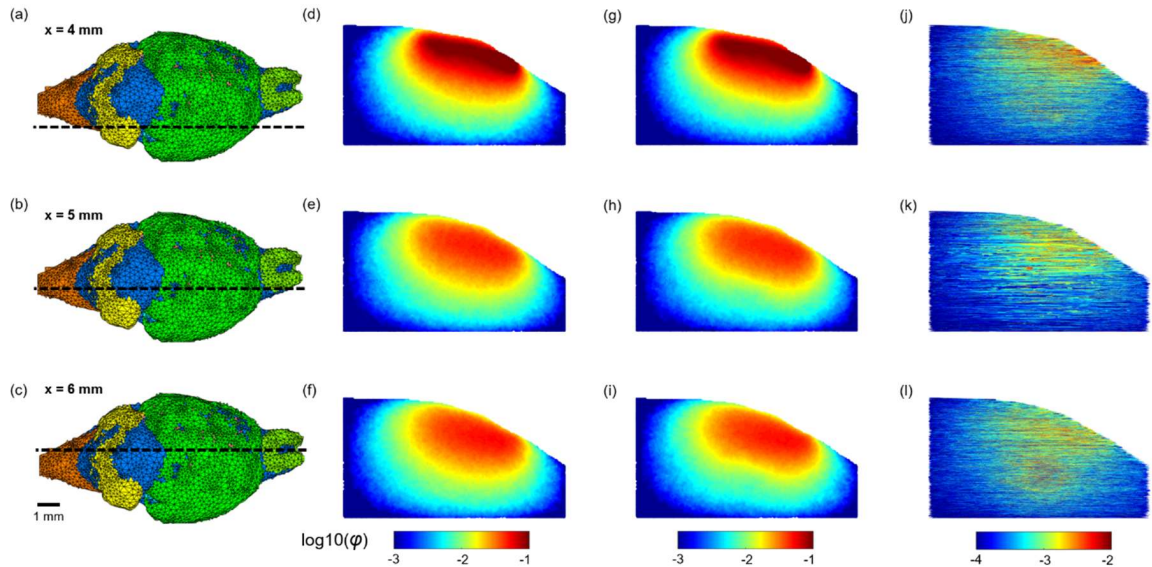

Figure 4. Optical fluence distribution maps of representative sagittal planes in 3D MC models with and without blood vessels. (a-c) Black dashed lines indicate the locations of the sagittal planes. (d-f) Normalized optical fluence maps without blood vessels shown in log scale. (g-i) Optical fluence maps with blood vessels. (j-l) Optical fluence difference between (d-f) and (g-i).

Although the optical fluence distributions on the global level are similar with and without blood vessels, the blood vessels introduce clear local fluctuations in the fluence distribution. The local fluctuations are expected to be more pronounced for actual *in vivo* applications, as the structural and functional conditions are much more complex. Both the global decay and local fluctuation in optical fluence contribute to the heterogeneity of the fluence distribution in PA imaging. Consequently, such heterogeneous fluence distribution causes errors in the quantification of the chromophore concentration. Most of the conventional PA reconstruction algorithms, such as the back-projection based method, do not take the fluence heterogeneity into account, resulting in low quantitative accuracy [53].

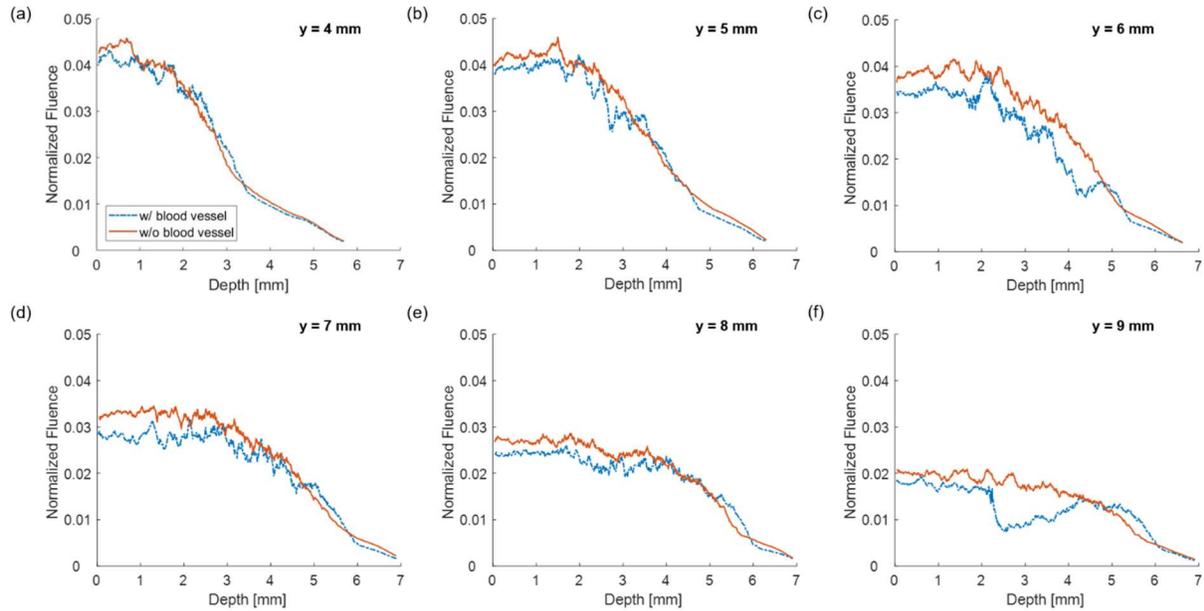

Figure 5: Normalized optical fluence as a function of depth, i.e. the distance from the skin surface. Six locations are selected along the sagittal suture (around x = 6 mm).

**3.2** *The influence of heterogeneous optical fluence on quantitative PACT*

The impact of heterogeneous optical fluence on quantitative PA image reconstruction is shown in Figure 6. Using k-wave, we simulated two different PA signal generation conditions with homogenous optical fluence and the heterogenous fluence (Figure 6(a-b)). A 7 × 5 mm$^2$ (y-axis ranges from 3 to 10 mm, z-axis ranges from 1 to 6 mm, as shown on Figure 2(b)) region was extracted from the fluence map at sagittal plane x = 6 mm. The initial PA pressure for the forward k-Wave simulation was normalized by the fluence map. The acquired acoustic signals were reconstructed by using the time-reversal based method (Figure 6(c-d)) and the delay-and-sum based method (Figure 6(e-f)), both of which are widely used for PA imaging reconstruction. Both methods assumed that the optical fluence was homogenous inside the brain. Under the homogenous fluence condition, the reconstruction assumption is valid, and thus the reconstructed PA image can better recover the actual $\mu_a$ of the target, despite the missing vertical structures due to the limited-view detection [54,55]. However, under the heterogenous fluence condition, which is more realistic for *in vivo* imaging, the reconstruction assumption is no longer valid. Targets in the deeper and peripheral regions receive lower optical fluence and thus generate weaker PA signals. As a result, the recovered $\mu_a$ of the weakly-excited targets is underestimated in the reconstructed image, as shown by the white arrows in Figures 6(c-d). Moreover, as tissue's optical properties are highly wavelength-dependent, the fluence map inside the tissue is also wavelength dependent, a common phenomenon in multispectral PA imaging also known as spectral coloring.

Omitting the heterogeneous fluence distribution is likely to yield an inaccurate estimation of chromophore concentrations in multispectral PA imaging and thus functional parameters such as blood oxygenation.

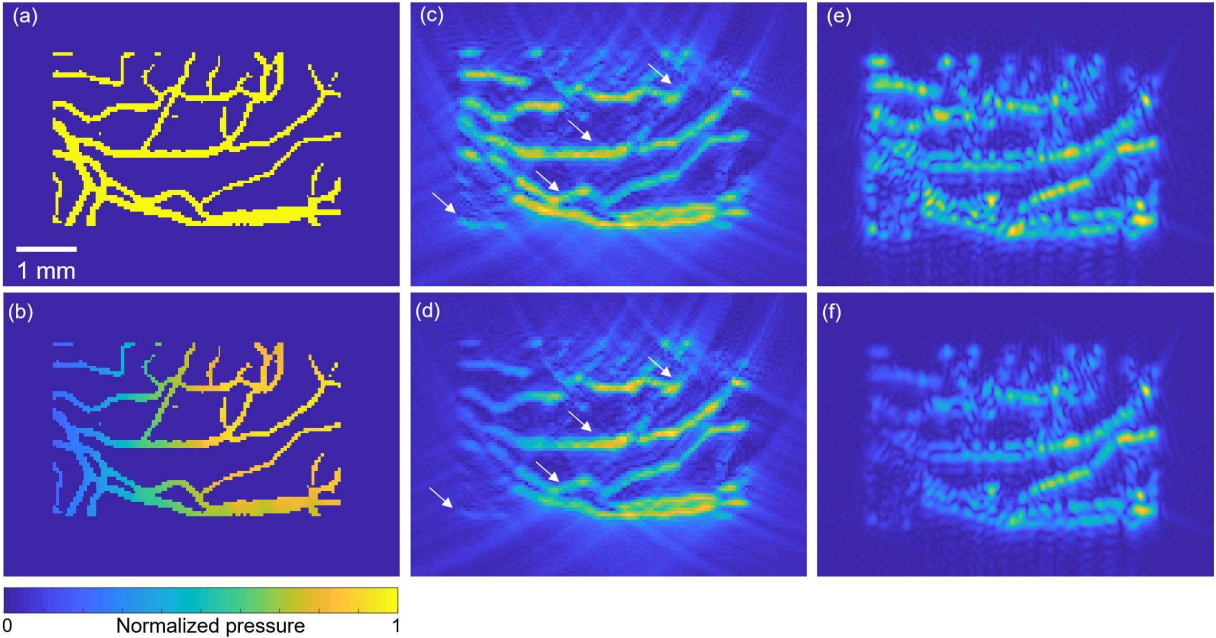

Figure 6. Simulated photoacoustic image reconstruction using k-wave. (a-b) Initial PA pressure map of blood vessels with homogenous and heterogenous fluence distribution, respectively. (c, d) Reconstructed blood vessel images by the time-reversal based method, using the initial pressure maps in (a-b), respectively. (e, f) Reconstructed blood vessel images by the delay-and-sum based method, using the initial pressure maps in (a-b), respectively.

## 4 Discussion & conclusion

With its balanced resolution and penetration depth and the inherent functional sensitivity, PACT is of high interest to the large neuroscience community. We have previous studied the skull's impact on the acoustic wave propagation in PACT of mouse brain [56]. In this paper, we investigate the impact of optical fluence distribution in mouse brain on quantitative PACT. Our results show that, mouse brain is a highly scattering organ, and the global decay of the optical fluence with depth is non-negligible regardless of the existence of blood vessels. The optical fluence decreases by a factor of 4-5 over the entire field of view. The existence of blood vessels introduces additional local fluctuations in optical fluence, due to optical property variation within the tissue structure. Both the global decay and the local fluctuations can lead to inaccurate estimation of chromophore concentrations in quantitative PACT.

Our 3D MC simulation accuracy depends on the digital mouse brain model, which was generated by combining the anatomical information and vascular information from two different data sets. It is technically challenging to register two data sets accurately. In this study, the two data sets were registered manually by mapping their functional regions to the maximum extent while ensuring that the final CBV ratio was consistent with the literature (~5%). To improve the model's accuracy, regional CBV ratios should also be considered. For example, as shown by Chugh *et al.*, cerebral cortex, hypothalamus, and hippocampus have a total CBV ratio of 7.9%, 4.5%, and 3.7%, respectively [42]. The regional differences in both CBV ratio and vasculature pattern are non-negligible. An adaptive registration should be developed to better match different regions of the brain and further improve the accuracy of the combined mouse brain model.

Though our mouse brain model included different functional regions, we used the optical properties extracted from human brain data, due to the lack of published values for moue brains at 1064 nm. Several studies have shown that the optical properties (especially $\mu_s$) vary for different functional regions of the mouse brain, mainly due to different composition of cell types [57–59]. Even though the differences might be insignificant, more accurate optical properties can further improve our mouse brain model's accuracy.

We used k-Wave simulation to demonstrate the impact of the global decay and local fluctuations in optical fluence on the reconstructed PA images. However, such simulation results are oversimplified. We used only 2d k-Wave simulation and did not consider out-of-plane signals or acoustic attenuation/reverberation. We did not consider the skull's impact either. Liang *et al.* have shown that the skull causes strong aberrations in high-frequency transcranial PA signals, leading to inaccurate target location and deteriorated PACT image quality [56]. Therefore, for *in vivo* applications, we expect that the optical fluence heterogeneity is much more complex, and its impact on quantitative PACT is more significant [56].

**Acknowledgement:** This work was supported in part by National Institute of Health (R01 EB028143, R01 NS111039, R01 NS115581, R01 GM134036, R21 EB027304, R43 CA243822, R43 CA239830, R44 HL138185); Duke MEDx Basic Science Grant; Duke Center for Genomic and Computational Biology Faculty Research Grant; Duke Institute of Brain Science Incubator Award; and American Heart Association Collaborative Sciences Award (18CSA34080277). The authors thank Caroline Connor for editing the manuscript.